# THE CLUSTERING OF QUASARS

P. Andreani, S. Cristiani & F. La Franca

*Dipartimento di Astronomia dell'Università, Padova, Italy.*

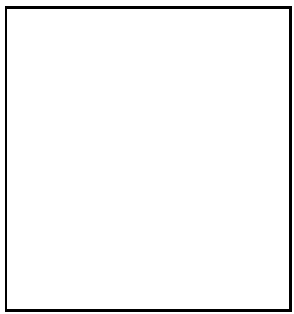

**Abstract**

We have investigated the spatial distribution of quasars and its relationship with redshift by using the two-point correlation function, the variance of cell counts and the conditional density as a function of redshift. By comparing our results with those found by other authors we conclude that quasars have a correlation function with amplitude similar to or greater than that of galaxies, but the issue of its evolution still needs to be settled. We discuss how this can be achieved by adding new observations and comparing real data with survey simulations.

## 1 Introduction

As widely discussed in this meeting the investigation of the large scale structure of the universe and its connection with the primordial density fluctuations, which in the gravitational instability picture gave origin to it, represents one of the major problems in modern cosmology. In order to tackle this, we have still to face two fundamental problems: on the one hand, we lack a theoretical picture relating the present day distribution of matter with the primordial fluctuations. There are actually first attempts to determine the evolution of the mass two-point correlation function from the linear regime to the present-day one ([14] [21], [19], [17] and Jain, this meeting), but the fundamental question of its relationship with the galaxy two-point correlation function needs still a deep understanding of the bias mechanism. On the other hand, we do not have any knowledge of the distribution of objects at high redshift since the available galaxy catalogs provide the determination of galaxy clustering only at relatively low redshifts. However, this problem can be soon challenged by studying quasars and their relationship with galaxies. From an observational point of view there are in fact two novelties: (1) the rapid growth of quasar surveys, which will help in assessing the statistical significance of QSO distribution and (2) the investigation of QSO's environments, which will relate QSOs with galaxies (see Boyle, this meeting).

## 2 The Quasar two-point correlation function

QSO clustering has been investigated by several authors and a detailed study of this effect at scales $r \leq 150\ h^{-1}$ Mpc ([24], [22], [1], [12] [2]) has been possible only with the availability of complete and faint quasar samples with *high* surface density, while previous attempts failed ([20]; [25], [7], [5], [11], [6]). The quasar two-point correlation function (CF) shows a shape and an amplitude, within the errorbars, similar to those of the galaxy CF: parametrizing the CF with the usual power-law model, $\xi(r) = (\frac{r}{r_0})^\gamma$, provides an amplitude $r_0$ in the range $6 \div 10\ h^{-1}$ Mpc. The same conclusion but with a higher S/N ratio is also found by using the conditional density, $\Gamma$, i.e. the mean density about a given object ([2]). This function is better suited for samples without a definite mean density and gives a less ambiguous and more robust description of the distribution.

## 3 The variance of Quasar counts

An independent way of establishing the clustering strength of objects is the measurement of the variance $\sigma^2$ of counts in cells. The variance is related to the spatial CF by a volume average and with respect to the CF yields a more reliable evaluation of the clustering amplitude at the scale of the cell–size.

The analysis has been carried out in the redshift range $1.0 \div 2.2$ for cells of volume $V_u$. A maximum likelihood analysis gives estimates of $\sigma^2(\ell)$ ($\ell \equiv V_u^{1/3}$) for $\ell = 40, 60, 80$ and $100\ h^{-1}$ Mpc: $\sigma^2(\ell) = 0.46^{+0.27}_{-0.27}$, $0.18^{+0.14}_{-0.15}$, $0.05^{+0.14}_{-0.05}$ and $0.12^{+0.13}_{-0.12}$, respectively, where the 70% confidence ranges account for both sampling errors and statistical fluctuations in the counts (see [4] for details). This allows a comparison of QSO clustering on large scales with analogous data obtained both for optical and IRAS galaxies: QSOs seem to be more clustered than these galaxies by a biasing factor $b_{QSO}/b_{gal} \sim 1.4 - 2.3$.

## 4 The evolution of QSO clustering

The increasing number of complete samples permits now to study not only the global quasar clustering with high statistical significance but also its cosmic evolution. At present seemingly contradictory results have been supported: Iovino et al. ([13]), Mo & Fang ([18]) and Komberg et al. ([15]) claim evidence for an evolving two-point correlation function with redshift in comoving coordinates, $\xi(z) \propto (1+z)^{-(2 \div 3)}$, while Andreani & Cristiani ([3]) and Shanks & Boyle ([23]) do not find such an effect and suggest that a stable clustering, $\xi(z) \sim const$, in comoving coordinates is consistent with the data. Probably this discrepancy can be reconciled once poor statistics and the partial knowledge of the selection effects affecting the samples will be better taken into account.

## 5 New data and the simulation of surveys

In order to address issues to the evolution of clustering we have pursued our investigation twofold: on the one hand we have enlarged our sample by including the deepest fields of the HBQS survey (*Homogenous Bright Quasar Survey*, [9]) covering a large connected area; on the other hand we have followed an inverse approach by building mock samples and comparing them with the real ones.

The mock surveys have been constructed as follows: (a) absolute magnitudes and redshifts are randomly extracted from a probability distribution function given by the adopted quasar luminosity function ([10], [16]). (b) The evolution with redshift is given by a simple pure luminosity evolution model: the luminosity function is assumed to translate towards brighter magnitudes with redshift with a law $M_B = M_B^\star - 2.5k\log(1+z)$. (c) Angular position of each object is found according to a clustering model, built by parametrizing the correlation function of quasars as usual, $\xi(r) = (\frac{r}{r_0})^{-\gamma}$, and assigning to each object a probability of being clustered found by a weighted volume average of this function. (d) The absolute magnitude is converted into apparent magnitude by taking account of the k-corrections (as in [8]) and to this photometric errors are randomly assigned. (e) Finally these mock catalogs are 'convolved' with the observation selection function given by the limiting fluxes, area, magnitudes, and selection criteria for candidates of each survey. We consider only objects with absolute B magnitude $M_B \leq -23$ in a flat universe ($\Omega = 1$) with H$_o$=50 km/s/Mpc and $q_o = 0.5$.

Results from 1000 simulations of each survey are added and the redshift distribution is then compared with the real one. A Kolmogorov-Smirnov test gives always a very high probability (always >0.20) that the two distributions are drawn from the same population. Finally we compute the number of pairs in shells of depth of 10 Mpc between 0 and 40 Mpc for the simulated surveys and the real ones. Errors on the simulated pairs are negligible, while those for the real ones are obtained with bootstrapping techniques. Figure 1 shows the number of pairs in the real case (filled squares) and those resulting from simulations. The real qso pairs undoubtedly lie above those expected from a poissonian distribution and exclude most of the selected models but those with $r_0 \sim 6 \div 10$ h$^{-1}$ Mpc. This approach is being used also for investigating the evolution with redshift. While issues to the evolution of qso-pairs can be properly addressed with the present data, their statistical significance is not enough to distinguish the most common hypotheses of the kind of clustering evolution (stable clustering, linear evolution, collapsing). To this aim we are completing a sample of 600 quasars selected over an area of 25 deg$^2$ with $B \leq 20.5$. This large connected and deep area will soon show unambigously the kind of evolution at the 95% confidence level. In fact, in the hypothesis of one of the milder clustering evolution in which $\xi(r) = \xi(r)(1+z)^{-1.2}$ (stable clustering, we would expect 67 pair at low redshift against 43 at high redshift.

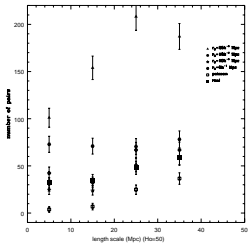

Figure 1: The qso pairs in shells of depth of 10 Mpc centred at 5, 15, 25 and 35 Mpc. Real data are shown as filled squares, the poissonian distribution as open squares. The other symbols refer to simulations corresponding to different clustering models, whose clustering amplitudes, $r_0$, are listed in the upper right corner.